\documentclass[useAMS,usenatbib]{mn2e} 
\usepackage{aas_macros}
\usepackage{graphics}
\usepackage[pdftex]{graphicx}
\usepackage{epstopdf}
\usepackage{epsfig}  
\usepackage{natbib} 
\usepackage{dingbat}
\usepackage{float}
\usepackage{amsmath}
\usepackage{times}
\usepackage[varg]{txfonts}
\usepackage{verbatim} 
\usepackage{multirow,bigdelim} 
\usepackage{color}
\usepackage{vmargin}
\setmarginsrb{1.2cm}{1cm}{1.2cm}{1cm}{1cm}{1cm}{1cm}{1cm}

  \makeatletter
    \renewcommand{\paragraph}{\@startsection{paragraph}{4}{\z@}%
      {-3.25ex\@plus -1ex \@minus -.2ex}%
      {1.5ex \@plus .2ex}%
      {\normalfont\small\centering}}
     
    \renewcommand{\subparagraph}{\@startsection{subparagraph}{5}{\z@}%
      {-3.25ex\@plus -1ex \@minus -.2ex}%
      {1.5ex \@plus .2ex}%
      {\normalfont\small\centering}}
    \makeatother

\newcommand{\kms}{{ km~s$^{-1}$}}
\newcommand{\hMpc}{{ \textit{h}$^{-1}$~Mpc}}

\title[Grouping]{How does the grouping scheme affect the Wiener Filter reconstruction of the local Universe?}
\author[Sorce \& Tempel]
{{Jenny G. Sorce$^{1,2}$\thanks{E-mail: \text{jenny.sorce@astro.unistra.fr / jsorce@aip.de}}, 
Elmo Tempel$^{2,3}$}\\
$^1$Universit\'e de Strasbourg, CNRS, Observatoire astronomique de Strasbourg, UMR 7550, F-67000 Strasbourg, France\\
$^2$Leibniz-Institut f\"{u}r Astrophysik, An der Sternwarte 16, 14482 Potsdam, Germany\\
$^3$Tartu Observatory, Observatooriumi 1, 61602 T\~oravere, Estonia
}

\begin{document}

\date{}

\pagerange{\pageref{firstpage}--\pageref{lastpage}} \pubyear{2017}

\maketitle

\label{firstpage}

\begin{abstract}
\indent 
High quality reconstructions of the three dimensional velocity and density fields of the local Universe are essential to study the local Large Scale Structure. In this paper, the Wiener Filter reconstruction technique is applied to galaxy radial peculiar velocity catalogs to understand how the Hubble constant (H$_0$) value and the grouping scheme affect the reconstructions. While H$_0$ is used to derive radial peculiar velocities from galaxy distance measurements and total velocities, the grouping scheme serves the purpose of removing non linear motions. Two different grouping schemes (based on the literature and a systematic algorithm) as well as five H$_0$ values ranging from 72 to 76~\kms~Mpc$^{-1}$ are selected. The Wiener Filter is applied to the resulting catalogs. Whatever grouping scheme is used, the larger H$_0$ is, the larger the infall onto the local Volume is. However, this conclusion has to be strongly mitigated: a bias minimization scheme applied to the catalogs after grouping  suppresses this effect. At fixed H$_0$, reconstructions obtained with catalogs grouped with the different schemes exhibit structures at the proper location in both cases but the latter are more contrasted in the less aggressive scheme case: having more constraints permits an infall from both sides onto the structures to reinforce their overdensity. Such findings highlight the importance of a balance between grouping to suppress non linear motions and preserving constraints to produce an infall onto structures expected to be large overdensities. Such an observation is promising to perform constrained simulations of the local Universe including its massive clusters.

\end{abstract}


\section{Introduction}

On large scales, where the gravity prevails, the Universe is homogeneous and isotropic enough for the observed velocity field to reflect the evolution of the Large Scale Structure (LSS) and the total underlying mass (i.e. both baryonic and dark) distribution. Therefore, to study the formation and evolution of the LSS, the analysis of observational radial peculiar velocities plays a major role \citep[e.g.][]{1994ARA&A..32..371D,1995PhR...261..271S,1999astro.ph..9003W,1999Obs...119..292D}. Consequently, several techniques have been developed to analyze the observed velocity datasets renewing the effort to measure them \cite[e.g.][]{1992ApJS...81..413M,1995MNRAS.276.1391N,1997ApJS..109..333W,1998MNRAS.299..425D,2001MNRAS.321..277C,2007ApJS..172..599S,2008ApJ...676..184T,2014arXiv1410.2992S,2016AJ....152...50T} and leading to numerous studies \citep[e.g.][]{1997ApJ...486...21Z,1998A&A...340...21T,1999ASPC..167..328T,2001MNRAS.326..375Z,2012ApJ...751L..30H,2012ApJ...744...43C,2013MNRAS.431.3678R,2015MNRAS.447..132W,2015MNRAS.449.4494H,2016MNRAS.461.4176H}. In particular, algorithms have been built to reconstruct from the sparse radial observational datasets, the three dimensional distribution of matter and the three dimensional velocity field \citep[e.g. POTENT, Wiener Filter, VIRBIUS, respectively][]{1999ApJ...522....1D,1999ApJ...520..413Z,2016MNRAS.457..172L}. Assuming a cosmological model as a prior, these methods are able to produce density and velocity fields of the local Universe on grids using for sole observational information the sparse and noisy radial peculiar velocity datasets. 

In this paper, we focus on the Wiener Filter (WF) algorithm \citep{1995ApJ...449..446Z}. This technique is very straightforward and Appendix A gives detailed equations. Briefly, based on correlation functions, derivation of matrices and their inverse, the Wiener Filter permits calculating readily the density and velocity fields assuming as a prior the power spectrum of a given cosmological model. While correlation functions are obtained with the power spectrum, the correlation vectors are derived with radial peculiar velocities called `the constraints'. These latter must be of high quality to allow exquisite reconstructions of the local Universe. Since the Wiener Filter is a linear minimal variance estimator, removing non linear motions in the observational catalogs seems primordial. A grouping scheme permits gathering galaxies that belong to a single cluster or group into one point. Subsequently, it produces one linear constraint (one position and radial peculiar velocity) against several non linear constraints that would damage the reconstruction obtained with the Wiener Filter. We thus seek to understand the impact of the chosen grouping scheme applied to the observational constraints on the resulting reconstructions. In addition, in view of the recent concerns and discrepancies regarding the Hubble constant value \citep[see][for a review]{2015LRR....18....2J}, we wish to study also the differences between reconstructions obtained with observational catalogs derived using different Hubble constant values. The Hubble constant permits indeed converting distances in~\hMpc\ units and most importantly it allows us to derive galaxy peculiar velocities from galaxy total velocities and distance measurements. Constraints are derived from the second sparse and noisy observational distance dataset of the Cosmicflows project\footnote{http://www.ipnl.in2p3.fr/projet/cosmicflows/} \citep{2013AJ....146...86T}.  

To summarize, this paper aims at determining the variance of the WF reconstruction with respect to the Hubble constant and grouping scheme choices. The final goal is to select the best choices to build constrained initial conditions of the local Universe within the CLUES\footnote{https://www.clues-project.org/} collaboration \citep{2010arXiv1005.2687G}. To study in detail our cosmic environment, the resulting performed simulations should resemble the local Universe down to the clusters. In particular, we expect to optimize the reproduction of the local massive clusters that have been slightly under massive so far, if not for the Virgo cluster \citep{2014MNRAS.437.3586S,2016MNRAS.460.2015S,2016MNRAS.455.2078S,2016MNRAS.460L...5C,2016MNRAS.458..900C}. These constrained simulations are the starting point of several projects to study the local Universe in detail, to understand our local environment and to compare it with observations.

This paper starts with a section describing the observational catalog of radial peculiar velocities or more precisely of galaxy direct distance measurements and the two grouping schemes compared here \citep[Tully private communication and][]{2016A&A...588A..14T}. In a subsequent section, the Wiener Filter algorithm is applied to the observational catalog grouped with the different schemes and applying different Hubble constant numerical values. First, the effect of the Hubble constant on the reconstructed velocity and overdensity fields are studied then, the impact of the grouping scheme on the reconstructed fields is analyzed in detail. An additional analysis made after minimizing the biases in the observational catalogs permits tempering the results. A conclusion presenting the best strategy for the next step (building constrained initial conditions) closes the paper. 


\section{Grouping \& Reconstruction techniques}

\begin{table*}
\begin{center} 
\begin{tabular}{cccccccc}
\hline\hline
(1) & (2) & (3) & (4) & (5) & (6) \\
Grouping Scheme & H$_0$ & $<$v$>$ & $\sigma_v$ & Skewness & Kurtosis & \\
+ Description &~\kms~Mpc$^{-1}$ & \kms & \kms & &  & \\
\hline
Tully: &72 &      17.2 & 1462 & -0.16 & 7.0 \\
4098 isolated, 910 groups &73 &   -74.9 & 1476 & -0.29 & 7.0 \\ 
Total: 5008 constraints &74 &  -166 & 1493 & -0.42 & 6.9 \\ 
444 groups with one distance measurement&75 & -258 & 1511 & -0.55 & 6.9 \\ 
On average 4.5 distance measurements per group&76 &   -350 & 1532 & -0.66 &  6.9 \\
\hline
Tempel: &72 &  59.5 & 1388 & -0.055 & 7.1 \\ 
3218 isolated, 2344 groups &73 &  -27.7 & 1401 & -0.20 & 7.0 \\
Total: 5562 constraints &74 &   -115 & 1417 & -0.34 & 6.9 \\%
1516 groups with one distance measurement&75 &  -202 & 1434 & -0.47 & 6.9 \\
On average 2.1 distance measurements per group&76 & -289 & 1454 & -0.60 & 6.8 \\
\hline
\hline
\end{tabular}
\end{center}
\vspace{-0.25cm}
\caption{Properties of the catalog of constraints (radial peculiar velocities) according to the grouping scheme: (1) Grouping scheme including a short description of the constituents of the catalog after grouping, (2) Hubble constant, (3) mean velocity, (4) standard deviation of the velocity distribution, (5) skewness of the distribution, (6) flatness of the distribution.}
\label{Tbl:0}
\end{table*}

\subsection{The Catalog}

The second generation catalog built by the Cosmicflows collaboration is a large publicly released catalog of radial peculiar velocities or more precisely of direct distance measurements.  Published in \citet{2013AJ....146...86T}, it contains more than 8,000 galaxy direct distance estimates. These measurements come mostly from the Tully-Fisher \citep{1977A&A....54..661T} and the Fundamental Plane \citep{2001MNRAS.321..277C} methods. Cepheids \citep{2001ApJ...553...47F}, Tip of the Red Giant Branch \citep{1993ApJ...417..553L}, Surface Brightness Fluctuation \citep{2001ApJ...546..681T}, supernovae of type Ia \citep{2007ApJ...659..122J} and other miscellaneous methods also contribute to this large dataset though to a minor extent ($\sim 12\%$ of the data). Using H$_0$=75.2~(=100h)~\kms~Mpc$^{-1}$ (the value given by \citeauthor{2013AJ....146...86T}, 2013), it extends up to about 250~\hMpc\ and about 50\% of the data are within 70~\hMpc\ and 90\% within 160~\hMpc. In a companion paper \citep{2017MNRAS.468.1812S}, we have shown that, in absence of a complete catalog and provided that it is properly grouped, the sampling of this catalog is optimal for Wiener Filter reconstructions with respect to uniformly distributed catalogs or catalogs of sole clusters. The goal is then to track the impact of the grouping technique on the resulting reconstructions. 

\subsection{The Grouping Schemes}

A grouped version designed by Tully, hereafter referred to as Tully Grouping Scheme, and released via the Extragalactic Distance Database\footnote{http://edd.ifa.hawaii.edu/} was used to build the first generation of constrained initial conditions that result in simulations resembling the local Universe down to 2-3~\hMpc\ \citep{2016MNRAS.455.2078S}. However, clusters reveal themselves to be under massive except for the Virgo cluster \citep{2016MNRAS.460.2015S} thanks to the prior minimization of biases introduced by \citet{2015MNRAS.450.2644S} that reduces the infall onto the local Volume, leads the monopole of the velocity field to zero and gaussianizes the distribution of observed radial peculiar velocities.

The difficulty resides in the definition of `group' itself. If on the simulation side, groups are well defined thanks to an access to the entire 3D information, on the observational side, calling an ensemble of galaxies a group constitutes a great challenge because of a restricted access to the information. In observations, knowing precisely the fraction of collapsed material becomes quite problematic. Still several schemes have been developed to define groups within galaxy catalogs. They mainly invoke Friends of Friends (FoF) like algorithms based on projected separation, radial velocities and even luminosities to identify what are called `groups' of galaxies \citep[e.g.][]{1982ApJ...257..423H,1983ApJS...52...61G,2002AJ....123.2976R,2004MNRAS.348..866E,2005MNRAS.357..608Y,2007ApJ...655..790C,2011MNRAS.412.2498M,2011MNRAS.416.2840L,2014MNRAS.441.1513O,2014A&A...566A...1T,2015MNRAS.449.1897O}. This paper does not aim at scrutinizing in detail the methods used to group catalogs. It aims at testing two recently released versions of groups for galaxies in the local Universe to understand the differences in the reconstructions generated by two various grouping schemes as described below:

\begin{itemize}
\item \citet{2016A&A...588A..14T} introduced a new grouping method (hereafter Tempel Grouping scheme). This method is based on a widely used FoF percolation method, where different linking lengths in radial (along the line of sight) and in transversal (in the plane of the sky) directions are used but the conventional FoF groups are refined using multimodality analysis. More precisely, \citet{2016A&A...588A..14T} use a model-based clustering analysis to check the multimodality of groups found by the FoF algorithm and they separate nearby/merging systems. In the current paper, we use published catalogs of groups detected using this new method.

\item Tully Grouping scheme is based on literature groups and in that respect is not a systematic scheme. Within 30~Mpc, groups are those identified by \citet{1987ApJ...321..280T}, further away groups are those given in the literature like Abell's catalog \citep{1989ApJS...70....1A}. Recently, \citet{2015AJ....149...54T,2015AJ....149..171T} published a more systematic way of deriving groups based on radii of second turn around and iterations. After comparisons, we find that the catalog grouped with this last scheme is an intermediate between the catalogs obtained with Tully and Tempel Grouping schemes and as such will result in more mitigated conclusions would we compare it to Tempel Grouping scheme. In addition, Tully Grouping scheme has been used so far with the second catalog to build constrained initial conditions. We thus stick to Tully Grouping scheme in the rest of the paper\footnote{Note that we reproduced the work with the 2015 Tully Grouping scheme and found that it gives as expected intermediate results between Tully and Tempel Grouping schemes.}.  
\end{itemize}

Tully and Tempel Grouping schemes provide the groups to which the different galaxies that populate the second catalog of Cosmicflows belong to as well as their total velocity (derived from the observed redshift). We note that the grouping schemes deliver groups built with a complete down to a magnitude limit sample of galaxies. Then, galaxies from the second catalog of Cosmicflows are distributed into these groups and only the groups to which they belong are retained for further use. The second catalog of Cosmicflows gives the individual distance modulus ($\mu$) measurements of each galaxy and their uncertainty ($\sigma_\mu$). To determine the radial peculiar velocity of the groups and their position in real space (by opposition with redshift space), we proceed as follows (Tully private communication):

\begin{equation}
\mathrm{\mu_{g}=\frac{\sum{w\times \mu}}{\sum{w}}} \ ; \ \mathrm{\sigma_{\mu g}=\sqrt{\frac{1}{\sum{\mathrm{w}}}}}
\ \mathrm{where\ w=\frac{1}{\sigma_{\mu}^2}},
\end{equation}

\begin{equation}
\mathrm{d_{g}=10^{\frac{\mu_{g}-25}{5}}} \ ; \ \mathrm{\sigma_{dg}=\sigma_{\mu g}\times \frac{log(10)}{5}},
\end{equation}

\begin{equation}
\mathrm{v_{pec\ g}= v_{tot\ g}- H_0\times d_g} \ ; \ \mathrm{\sigma_{vpec\ g}=\sigma_{dg}\times d_g\times H_0},
\end{equation}

\noindent where the subscript `g' stands for `grouped' value and $\sigma$ for the uncertainty of the given subscript value, d is the distance in real space, v$_{\mathrm{tot}}$ is the total velocity of the galaxy/group and v$_{\mathrm{pec}}$ is the radial peculiar velocity.

Table~\ref{Tbl:0} reflects the resulting grouped catalogs after application of the two schemes. The first column shows interestingly that while Tully scheme results in more isolated galaxies (i.e. single position and peculiar velocity as constraint for the Wiener Filter algorithm), Tempel scheme gives less isolated galaxies but more groups (2344 against 910). Overall, Tully scheme is more aggressive than Tempel scheme. While on average, there is 4.5 distance measurements per group with Tully scheme, there is on average only 2.1 distance measurements per group with Tempel scheme. However, this difference could be due to the absence of group identification in Tully scheme when there is only one galaxy measurement.  Indeed, summing in both cases the number of isolated galaxies and that of groups with only one measurement, the numbers become similar (4542 for Tully versus 4734 for Tempel). However, excluding groups with a single measurement, there is still on average more distance measurements per group with Tully scheme (7.8) than with Tempel scheme (4.1) confirming that Tully scheme groups more (number of groups with more than one measurement about twice smaller). In total, Tully scheme provides 5008 constraints against 5562 for Tempel scheme.

Table~\ref{Tbl:0} also gives the properties of the resulting radial peculiar velocity distributions according to both the grouping scheme and H$_0$ ranging from 72 to 76~\kms~Mpc$^{-1}$. The larger H$_0$ the smaller and more negative the mean velocity, the larger the standard deviation, the more asymmetric and less flat the distribution for both grouping scheme. However, whatever H$_0$ value considered, Tempel scheme results in smaller mean, standard deviation and skewness values. Note how the mean trend changes for a H$_0$ value of between 72 and 73~\kms~Mpc$^{-1}$ in both cases.

\subsection{The Wiener Filter Technique}

We apply the Wiener Filter technique to the 10 catalogs obtained with the 5 different H$_0$ values and the two grouping schemes using Planck power spectrum \citep{2014A&A...571A..16P} as a prior. One might argue that using a different H$_0$ value to build the catalog of constraints and the cosmological prior could bias the results. Note that tests we made changing the prior (for instance using WMAP7 instead of Planck power spectrum) show that the prior has only a very small, thus negligible, impact on the reconstruction with respect to the parameters (grouping scheme and H$_0$ in the observational data) tested in this paper. In other words, the variance between reconstructions obtained with different priors with all the other parameters fixed is much smaller  than the variance between reconstructions produced with the same prior but changing H$_0$ or the grouping scheme. 

A boxsize of 500~\hMpc\ is retained as the adequate size to contain all the data-constraints. Note that from now on, the discussion will be led in \hMpc. Namely, once H$_0$ has been chosen, every distance is converted in \hMpc\ such that H$_0$=100h~\kms~Mpc$^{-1}$. A grid size of 256$^3$ cells permits a resolution about 2~\hMpc, the linear theory threshold, in agreement with the maximum resolution of the linear WF method. This ensures that differences observed between reconstructions are solely due to the tested parameters and not to non linear statistical fluctuations.

Additionally, non linear sigmas, explained in more detail in Appendix A, are essential to account for the residual of non linearities in the datasets. Indeed, even grouped catalogs still contain non linearities especially in high density regions with a poor sampling. The non linear sigmas correspond to a small additional smoothing applied to the constraints to compensate for their non linear component that cannot be accounted for directly by the linear Wiener Filter technique. They are simply added in quadrature to the uncertainties of the constraints. Non linear sigmas of the same order of magnitude (100-200 \kms) are found to be required for the different catalogs. Such similar values will prevent any difference due to a significant change in the smoothing. These non linear sigmas are essential to ensure that only significant differences remained visible between reconstructions obtained with various parameters.

\section{WF Reconstructions}

\begin{table*}
\begin{center} 
\begin{tabular}{cccccccc}
\hline
\hline
(1) & (2) & (3) & (4) & (5) & (6) &(7)\\
Grouping Scheme & H$_0$ & $\sigma_v$ & $\sigma_\rho$ & Dipole at r=10~\hMpc & Dipole at r=240~\hMpc & Monopole at r=240~\hMpc\\
& \kms Mpc$^{-1}$ & \kms &  & \kms & \kms & \kms\\
\hline
        &        72	& 320 & 0.20 & 477 &  144 & -83  \\
        &        73  & 332 & 0.21 & 476 & 144 & -359 \\
        Tully &        74  & 369 & 0.21 & 478 & 144 & -634\\
         &        75  & 424 & 0.22 & 477 & 144 & -914\\
         &        76  & 491 & 0.23  &479 & 144 & -1192\\
         \hline
            &    72  & 324 & 0.22 &476  & 138  & -62\\
            &    73  & 339 & 0.22 & 445 & 138  & -264 \\ 
Tempel &    74  & 369 & 0.22 &478  & 140 & -611\\
            &    75  & 615 & 0.23 & 624 &  140 & -1201\\
            &    76  &495 & 0.24 & 420 &  183 & -797\\
\hline
\hline
\end{tabular}
\end{center}
\vspace{-0.25cm}
\caption{Properties of the reconstructed velocity and overdensity fields for different H$_0$ values and grouping schemes: (1) Grouping scheme, (2) Hubble constant, (3) standard deviation of the velocity field, (4) standard deviation of the overdensity field, (5) dipole value of the velocity field at 10~\hMpc, (6) dipole value of the velocity field at 240~\hMpc, the edge of the box/data, (7) monopole value of the velocity field at 240~\hMpc.}
\label{Tbl:1}
\end{table*}

\subsection{Tully Grouping: the results}

\begin{figure*}
\vspace{-0.5cm}
\includegraphics[width=0.515\textwidth]{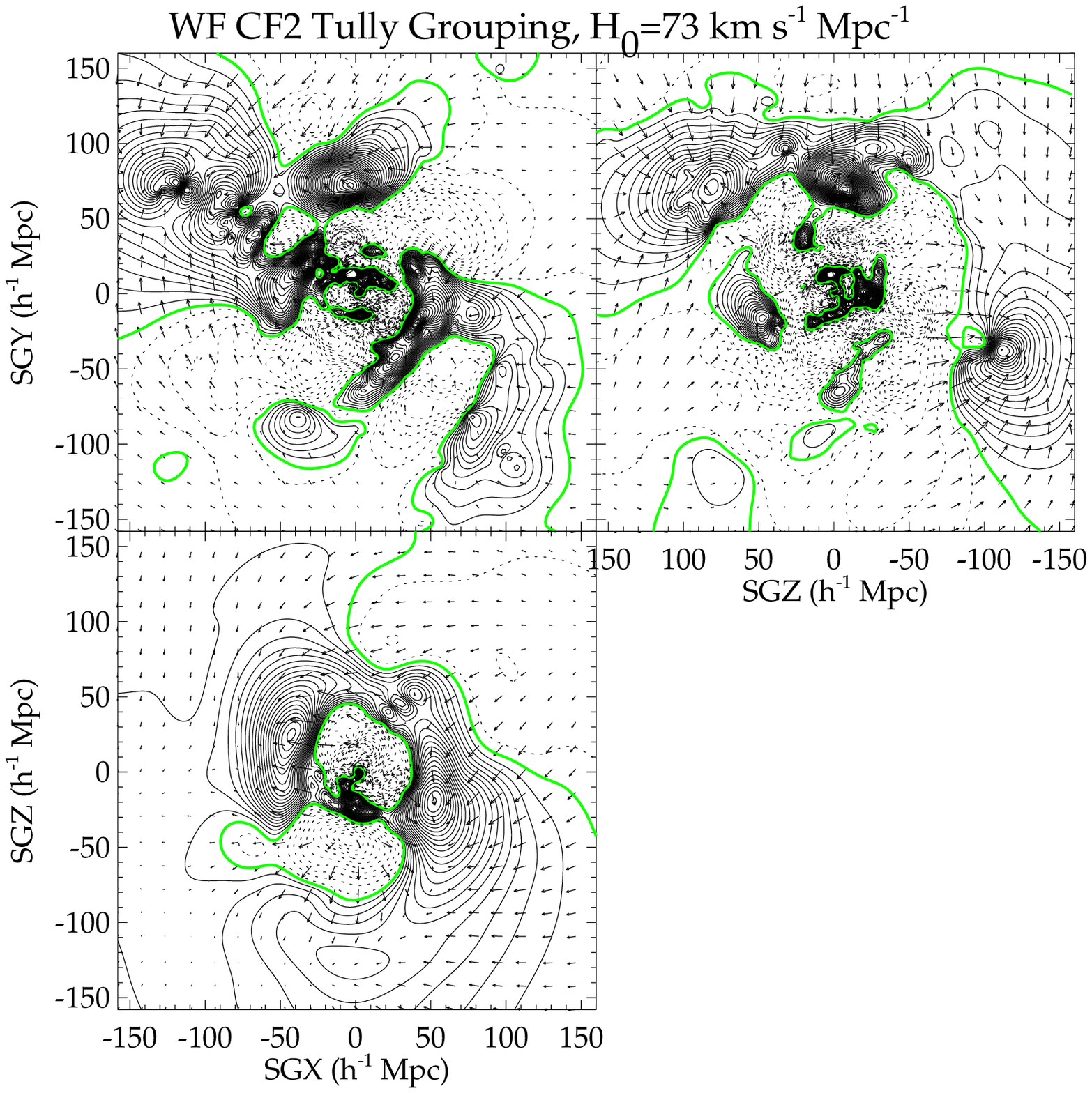}\hspace{-0.63cm}
\includegraphics[width=0.515\textwidth]{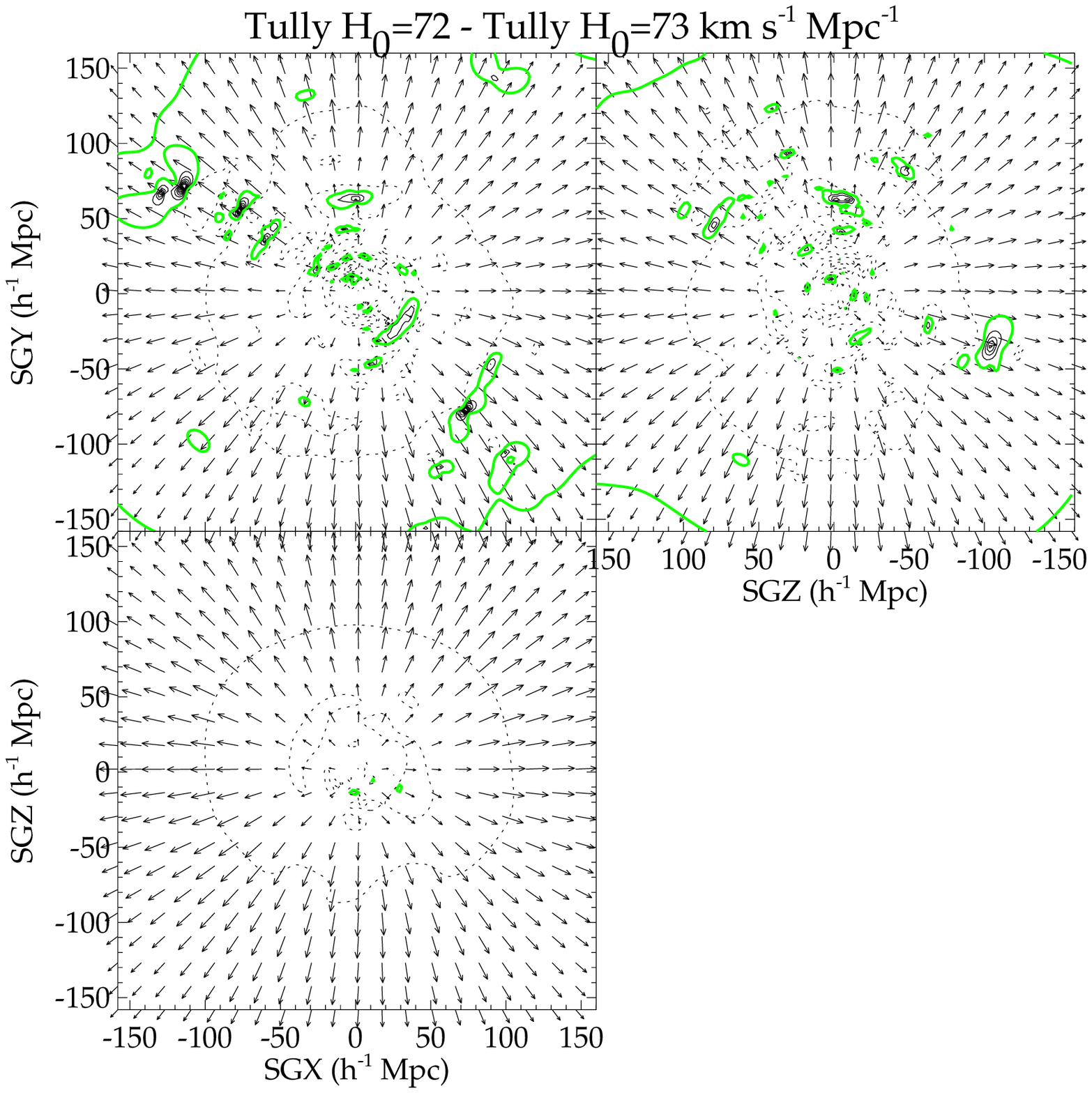}
\caption{Left: Supergalactic XY, YZ and XZ slices of the reconstructed velocity (arrow) and overdensity (contour) fields of the local Universe obtained with the catalog grouped with Tully scheme and H$_0$=73~\kms~Mpc$^{-1}$. The green color stands for the mean field. Dashed contours are underdense regions while solid contours are overdense areas. The reconstruction shows overall the local structures such as Shapley (top left in XY), Coma (top middle in XY and ZY) and Perseus Pisces (bottom right in XY). Right: as left panel but for the residual between reconstructions obtained with Tully grouping scheme and H$_0$=72 and 73~\kms~Mpc$^{-1}$ respectively. The residual highlights the impact of the Hubble constant value chosen to derive the distances and thus the peculiar velocity constraints. The larger H$_0$ is, the greater the infall onto the local Volume is.}
\label{fig:tullyfield}
\end{figure*}

The left panel of Figure \ref{fig:tullyfield} shows the reconstructed velocity and overdensity fields obtained with the catalog grouped with Tully scheme and using H$_0$=73~\kms~Mpc$^{-1}$. The right panel of the same figure presents the residual between the reconstructed fields obtained with two different H$_0$ values but the same grouping scheme. The effect is clear, the larger H$_0$ is, the greater the infall onto the local volume is. Namely, H$_0$ value impacts the tidal part of the velocity field\footnote{The velocity field can be decomposed into two components, the tidal part due to the objects outside of the volume considered and the divergent part generated by the objects within the volume considered.}. However, the overdensity field is not that much affected: there are only very small and sparse residual contours. It means that H$_0$ value influences only weakly the divergent part of the velocity field directly linked to the overdensity field. Note that this is the part of the velocity field used to build constrained initial conditions. The infall observed with larger value of H$_0$ impacts the global density of the local Volume. With a smaller value of H$_0$, not only the infall but also the global local density decrease: the spherical dashed contours on Figure \ref{fig:tullyfield} right indicate indeed that globally the reconstructed field obtained with the highest value of H$_0$ has higher overdensity values than that obtained with the smallest value of H$_0$. Since an underdensity of the local Volume is not excluded \citep[e.g.][]{2013ApJ...775...62K} while a large infall onto the local Volume is very unlikely, the smallest values of H$_0$ tested here might be preferred. However, in the last part of this section, we will temper this conclusion by applying the bias minimization scheme introduced in Section 2 and that needs to be applied to the observational catalog.
\begin{figure}
\vspace{-2.5cm}
\hspace{-0.5cm}\includegraphics[width=0.58 \textwidth]{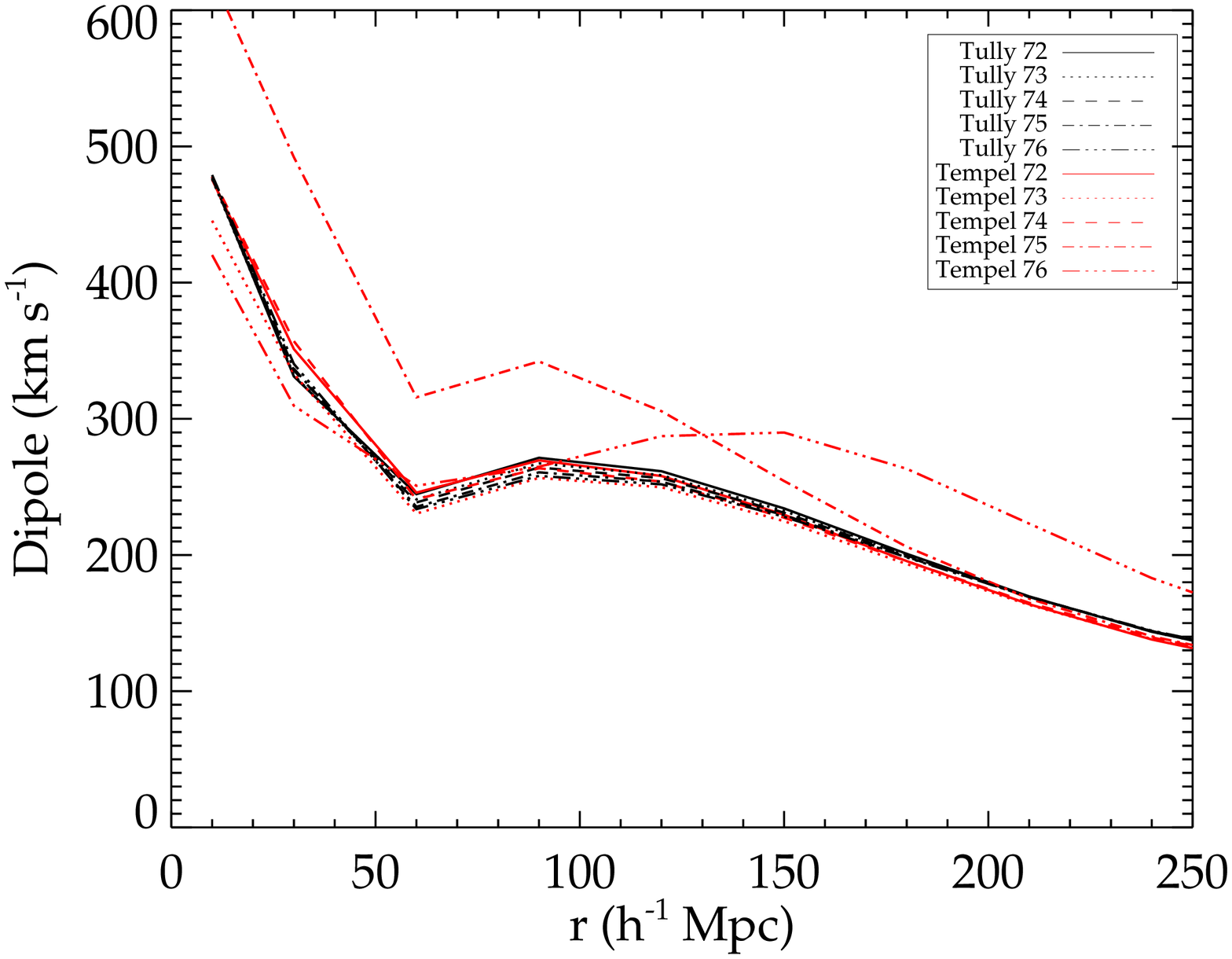} \\
\vspace{-4.cm}

\hspace{-0.5cm} \includegraphics[width=0.58 \textwidth]{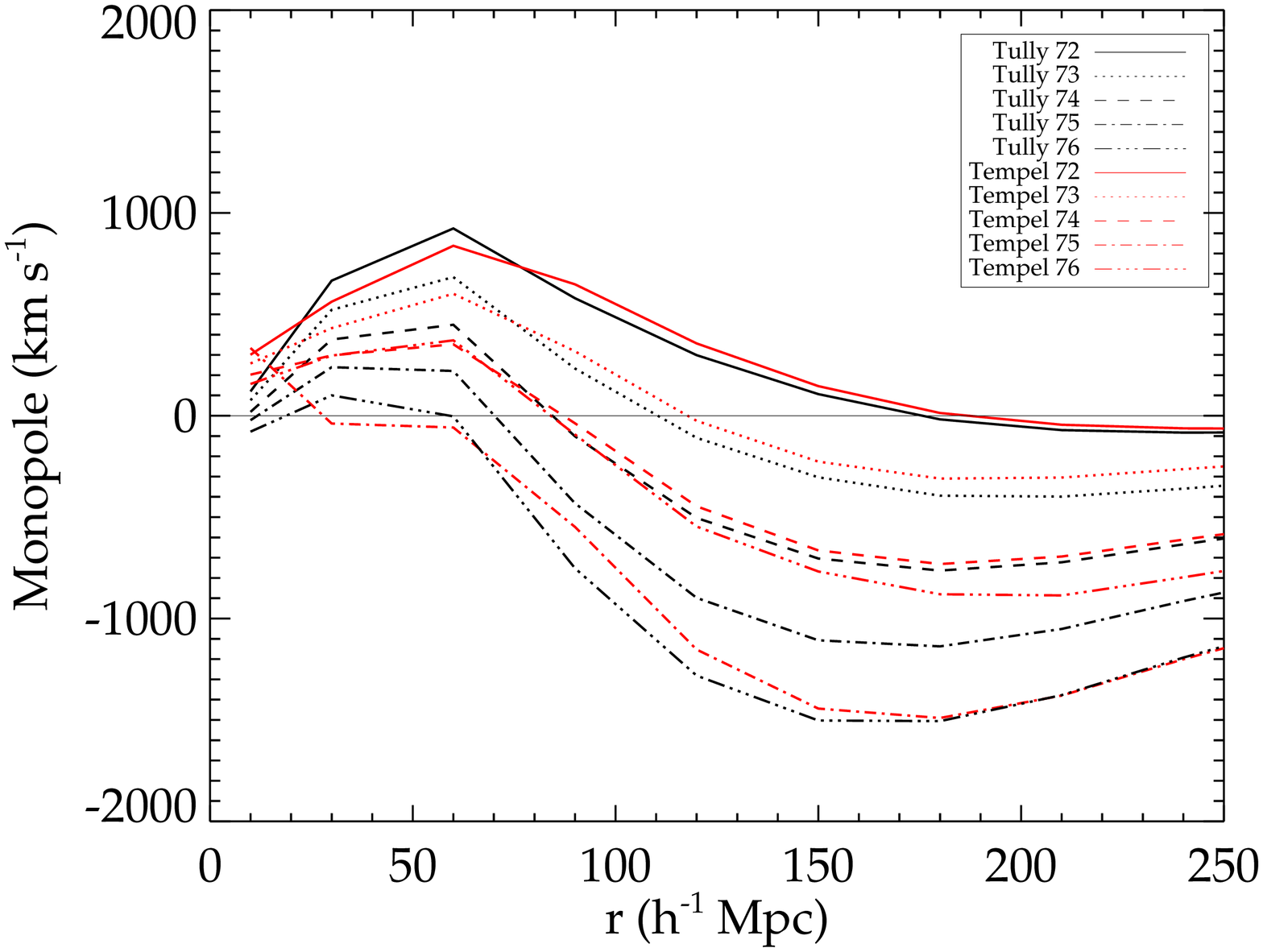}
 \vspace{-1.7cm}

\caption{Dipole (top) and monopole (bottom) of the Wiener Filter reconstructed fields for different H$_0$ values (linestyle) and different grouping schemes (color) as a function of the distance from us. If H$_0$ impacts only weekly the dipole, the monopole term is profoundly affected. The larger H$_0$, the larger is the infall on the local Volume. The grouping scheme has only a weak influence on the dipole and monopole of the velocity field except very locally. }
\label{fig:dipomono}
\end{figure}

The first half of Table~\ref{Tbl:1} summarizes the properties of the reconstructions obtained with Tully Grouping scheme and different H$_0$ values to support our findings based on Figure \ref{fig:tullyfield}. On the one hand, it clearly shows that for large H$_0$ values, the infall is large: the monopole term of the velocity field is highly negative at large radii. The infall for larger H$_0$values, deduced from the observed outflow in the subtraction of reconstructions obtained with increasing values of H$_0$ in the right panel of Figure \ref{fig:tullyfield}, is confirmed. At both large and small radii, the dipole of the velocity field is on the other hand quasi-unchanged, in agreement with the fact that the overdensity (or divergent part of the velocity) field is quite unaffected by a change in H$_0$ value. These two points are visible in another form on Figure \ref{fig:dipomono} where both monopole and dipole of the velocity field are shown at all radii. While the dipole is quite independent of H$_0$ value at all radii, the monopole tends to get smaller and smaller at all radii with H$_0$ getting larger and larger. On another aspect, the standard deviation of the overdensity and velocity fields increase slightly with the value of H$_0$. 

While Table~\ref{Tbl:1} shows properties of reconstructions obtained with different H$_0$ values independently of each other, the first third of Table~\ref{Tbl:2} summarizes the comparisons between reconstructed fields obtained with different H$_0$ values but the same (Tully) grouping scheme. Standard deviation of the residual between two different H$_0$ reconstructed overdensity and velocity fields obviously increase with the difference between the two H$_0$ values but are quite stable for a given difference between the two H$_0$ values. In any case, the standard deviation of the residual is smaller than the standard deviation of the compared velocity and overdensity fields taken independently except when the reconstructed velocity field obtained with 76~\kms~Mpc$^{-1}$ is compared to that obtained with the smallest H$_0$ value (i.e. 72~\kms~Mpc$^{-1}$), namely when the separation between H$_0$ values, chosen for this paper,  is maximal.  Regardless, 76~\kms~Mpc$^{-1}$ seems to be a very unlikely value in light of the above observations.

\begin{table}
\begin{center} 
\begin{tabular}{ccccc}
\hline
\hline
(1) & (2) &(3) & (4) & (5)  \\
Grouping & H$_0 $ 1- 2 & $\sigma_v$ & $\sigma_\rho$ & $\rho_{\mathrm{max}}$ \\
Scheme 1-2 & ~\kms~Mpc$^{-1}$& \kms &  & \\
\hline
  &   72 - 73 & 95 & 0.03 & 1.0 \\\
 &   72 - 74 & 187 & 0.06 & 1.4\\%
 &   72 - 75 & 282 & 0.1 & 1.5\\
   &   72 - 76  & 375 & 0.13 & 2.1 \\
Tully - Tully & 73 - 74 & 93 & 0.03 & 0.9 \\%
  &  73 - 75 & 187 & 0.06 & 1.2 \\%
 &   73 - 76 & 280 & 0.1 & 2.1\\%
 &   74 - 75& 94 & 0.03 & 0.9 \\%
 & 74 - 76& 187 & 0.06 & 1.3 \\%
 &  75 - 76& 93 & 0.03 & 1.3  \\%
 \hline
  &72 - 73  & 128 & 0.03 & 1.0\\%
 &   72 - 74 &187 & 0.06 & 1.4  \\%
 &   72 - 75 &448 & 0.1 & 2.0\\
  &   72 - 76 &427 & 0.1 & 2.4 \\%
Tempel - Tempel & 73 - 74 & 147 & 0.03 & 0.8\\%
  &  73 - 75 & 401 & 0.06 & 1.3\\%
 &   73 - 76 & 383 & 0.1 &1.8\\%
 &   74 - 75 & 298 & 0.03 & 0.9\\
 &  74 - 76 & 366 & 0.06 & 1.1\\%
 &   75 - 76 & 519 & 0.03 & 0.8\\%
 \hline
& 72 - 72  & 43 & 0.07 & 4.0  \\%
& 73 - 73  & 111 & 0.07 & 4.1 \\%
Tully - Tempel & 74 - 74 &43 & 0.07 & 4.3 \\%
& 75 - 75  & 254 & 0.07 & 4.2  \\%
& 76 - 76  & 400 & 0.07 & 4.7\\%
 
\hline
\hline
\end{tabular}
\end{center}
\vspace{-0.25cm}
\caption{Properties of the residual between reconstructed velocity and overdensity fields obtained with different H$_0$ values and different grouping schemes: (1) Grouping scheme of the reconstruction number 1 - Grouping scheme of the reconstruction number 2, (2) H$_0$ value used for the first reconstructed field - H$_0$ value used for the second reconstructed field, (3) standard deviation of the residual velocity field, (4) standard deviation of the residual overdensity field, (5) maximum of the residual overdensity field.}
\label{Tbl:2}
\end{table}


\subsection{Tempel Grouping: the results}
\begin{figure*}
\vspace{-0.5cm}
\includegraphics[width=0.515 \textwidth]{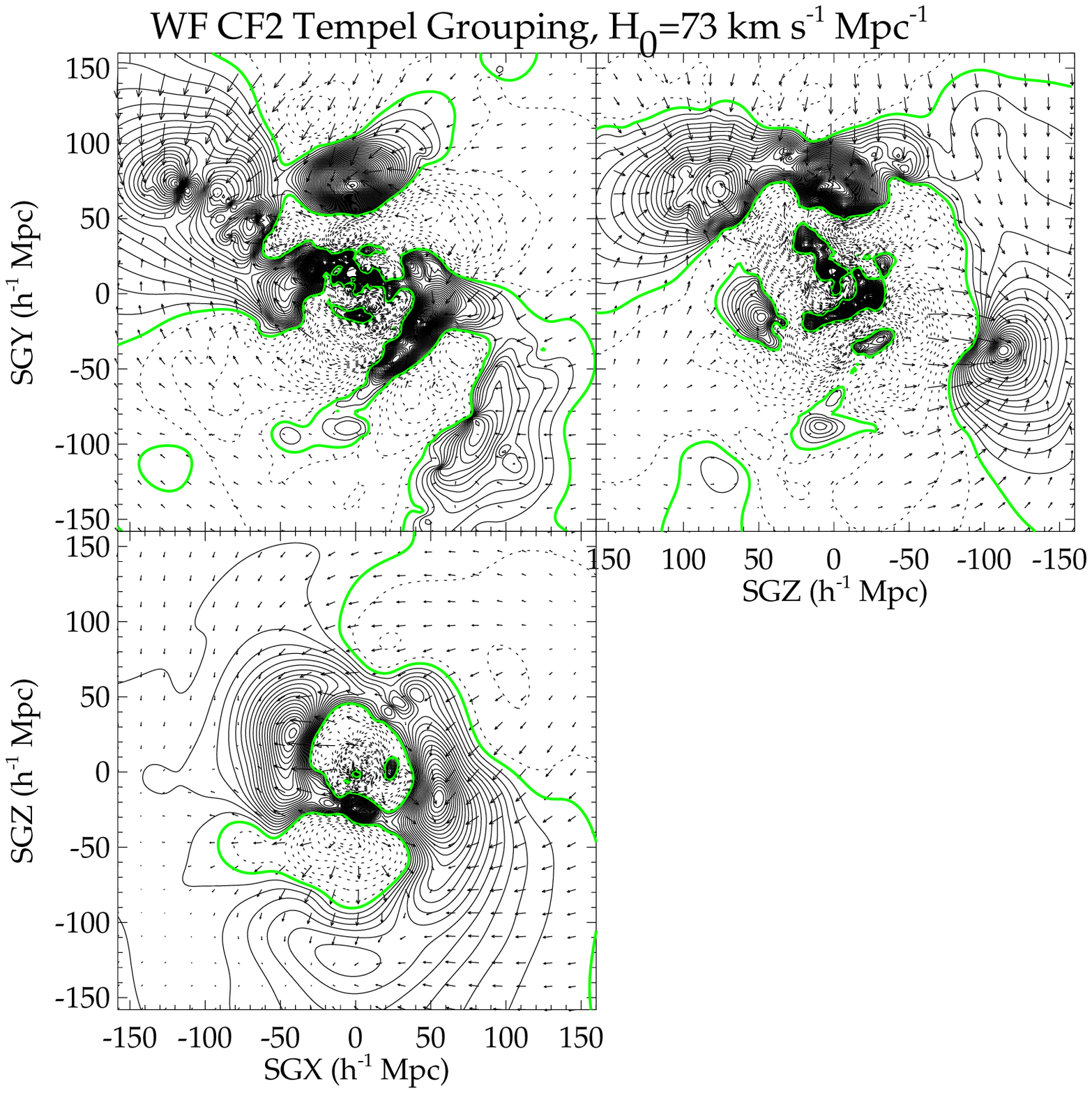}\hspace{-0.63cm}
\includegraphics[width=0.515 \textwidth]{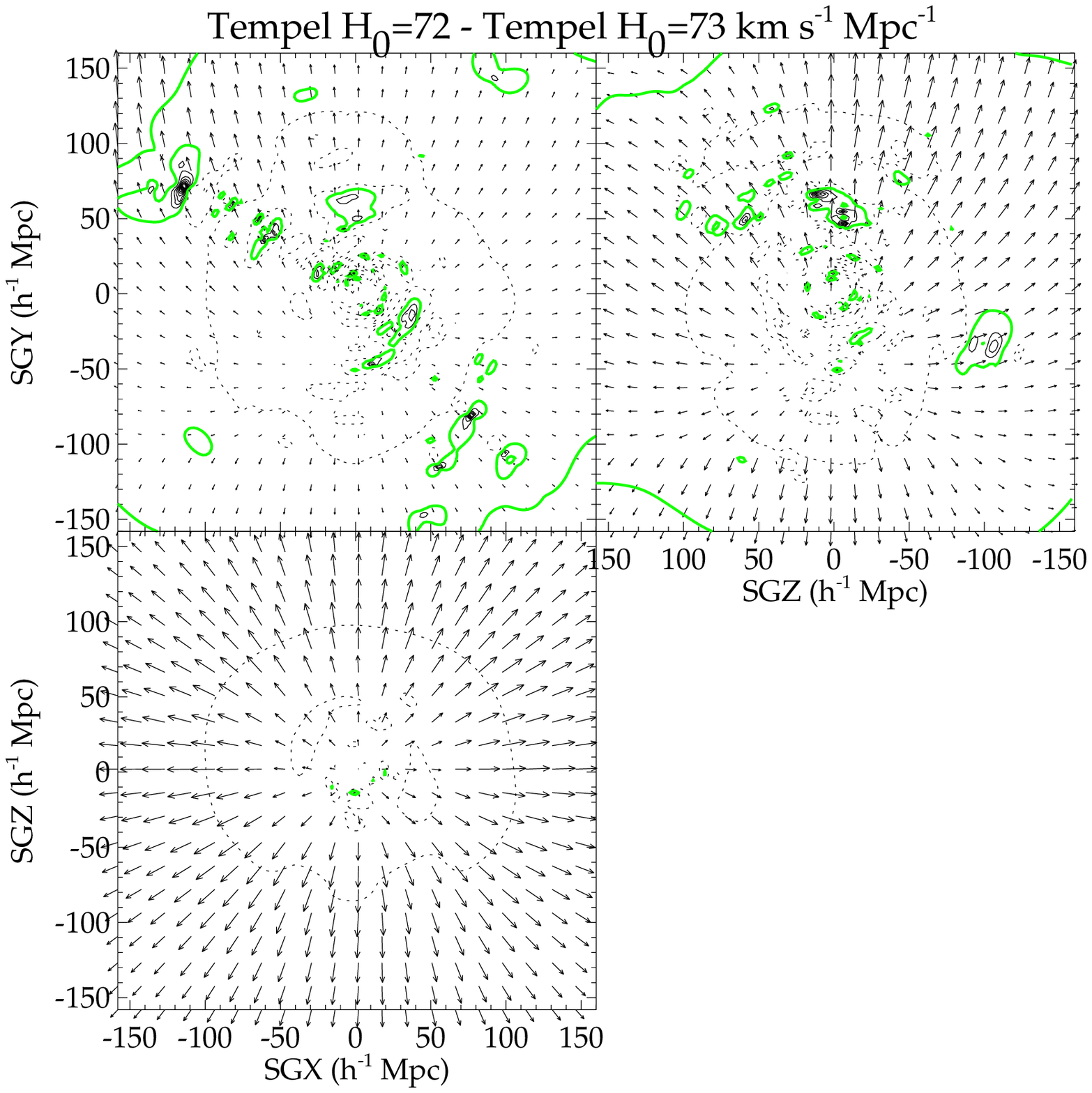}
\caption{As Figure \ref{fig:tullyfield} but obtained with Tempel grouping scheme.}
\label{fig:tempelfield}
\end{figure*}

Figure \ref{fig:tempelfield} shows the reconstructed velocity and overdensity field obtained with  H$_0$=73~\kms~Mpc$^{-1}$, like in Figure \ref{fig:tullyfield}, but with Tempel Grouping scheme. The observations made in the previous subsection still stand. Namely, H$_0$ value impacts clearly the tidal part of the velocity field while it barely affects the overdensity or divergent part of the velocity field. As H$_0$ gets larger, the infall onto the local Volume increases.

The second part of Table~\ref{Tbl:1} summarizes the different values obtained for the reconstructions obtained with Tempel Grouping scheme and different H$_0$ values. Again, the same findings as with Tully Grouping are valid except that the standard deviations of both the velocity and overdensity fields are slightly higher, a first hint that structures are more contrasted in the WF reconstructions obtained with Tempel Grouping. Tempel Grouping reconstructions are also less affected by the infall or in other words for a given H$_0$ value, the monopole term is less negative in the reconstructions obtained with Tempel Grouping than with Tully Grouping. The dipole varies slightly more in Tempel grouping scheme's case than in Tully grouping scheme's case probably because of the higher number of constraints: at small radii the larger number of constraints generates more non linearities, at large radii the larger number of constraint slows the fields in their pace to reach the mean value. The two largest values of H$_0$ (75 and 76~\kms~Mpc$^{-1}$) present exceptions that deserve attention. A value of 75~\kms~Mpc$^{-1}$ results in a larger dipole value than the average at small radii while a value of 76~\kms~Mpc$^{-1}$ gives a field with a larger dipole value than the average at large radii. In addition, the monopole value at large radii for H$_0$=75~\kms~Mpc$^{-1}$ is extremely high in absolute value. It clearly looks like there is a transition between values of 74 and 75~\kms~Mpc$^{-1}$ linked to the grouping scheme since none of these observations are valid for Tully Grouping scheme. This seems to imply that a more aggressive grouping has to be preferred for a better stability of the dipole and monopole of the velocity field whatever H$_0$ value is used. 

Tests we made varying the default linking length  (0.25 \hMpc\ at redshift zero changed to 0.20 or 0.30 \hMpc) in Tempel Grouping scheme and applying the WF technique to the resulting grouped catalogs show that indeed a large linking length permits increasing the stability but an excessive grouping (no more field galaxies) leads to wrong dipole values. This is in agreement with \citet{2017MNRAS.468.1812S} that show that galaxies in the fields are an absolute necessity. Additionally H$_0$ has to be chosen with more care: minimizing in absolute value the mean of the velocity distribution seems a reasonable approach to choose the value of H$_0$. Again, we will temper this conclusion within the last part of this section.

The second third of Table~\ref{Tbl:2} shows the properties of the residual between reconstructions obtained with Tempel Grouping scheme but different H$_0$ values. Overall, the same observations as with Tully Grouping scheme apply. One might notice that the residual values are larger than those obtained with Tully Grouping scheme. This is again due to the aggressiveness of the grouping. Indeed, in the tests made varying the default linking length, we observe that the variance between the two reconstructions obtained with different H$_0$ values is larger for the smallest linking length than for the default linking length used in the tests. Namely, grouping more eases slightly the dependence on H$_0$.

\subsection{Comparisons between the grouping schemes}

\begin{figure}
\vspace{-0.5cm}
\hspace{-0.63cm}\includegraphics[width=0.515 \textwidth]{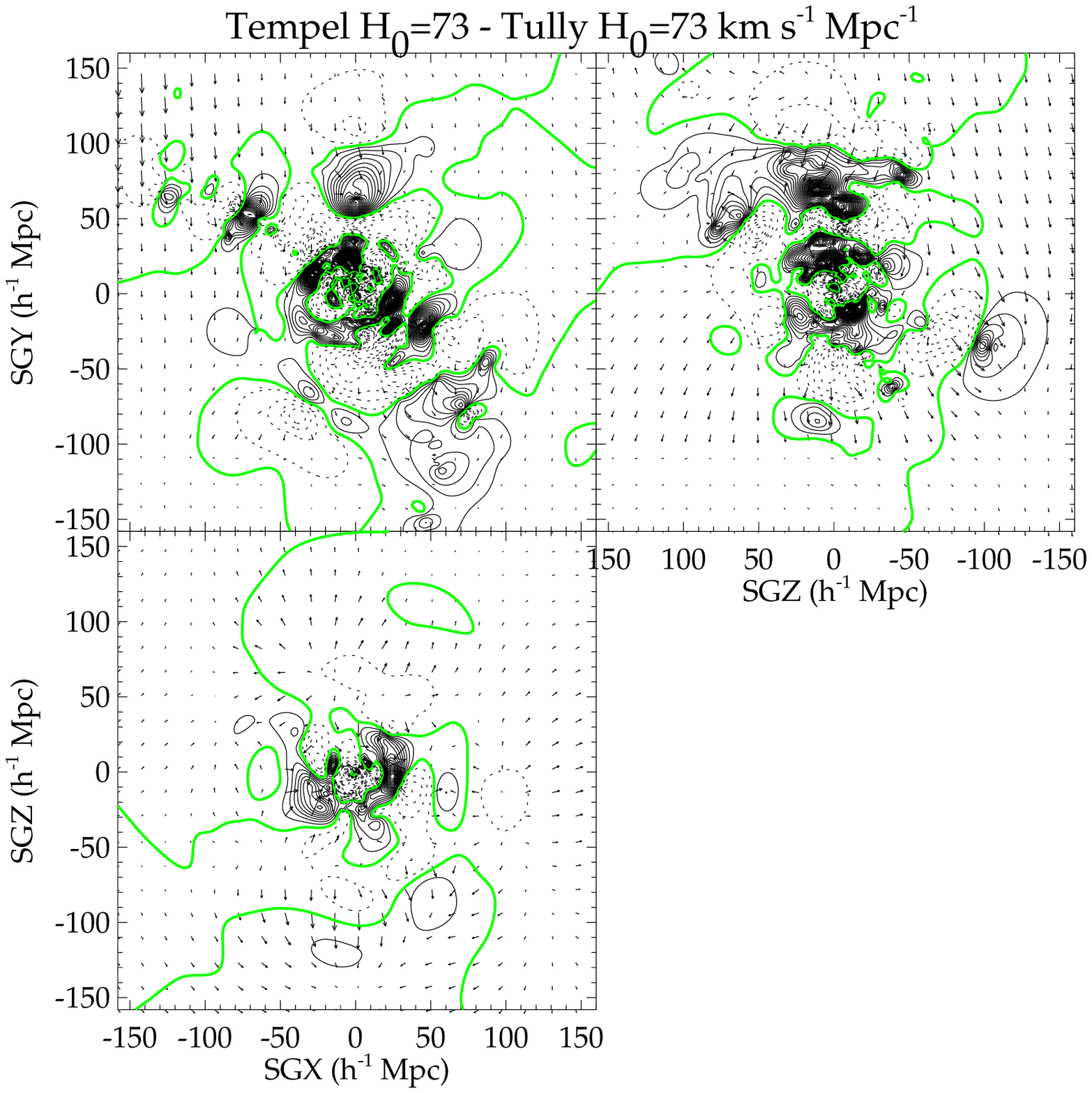}
\caption{Supergalactic XY, YZ and XZ slices of the residual between reconstructed velocity (arrow) and overdensity (contour) fields of the local Universe obtained with the catalog grouped with Tempel and Tully schemes. The green color stand for the null value. The residual shows that overall the local structures such as Coma (top middle in XY and ZY) are more pronounced in the reconstruction obtained with Tempel grouping scheme than in that obtained with Tully grouping scheme.}
\label{fig:tempeltullyfield}
\end{figure}

\begin{figure}
\vspace{-0.5cm}
\hspace{-0.8cm}\includegraphics[width=0.55 \textwidth]{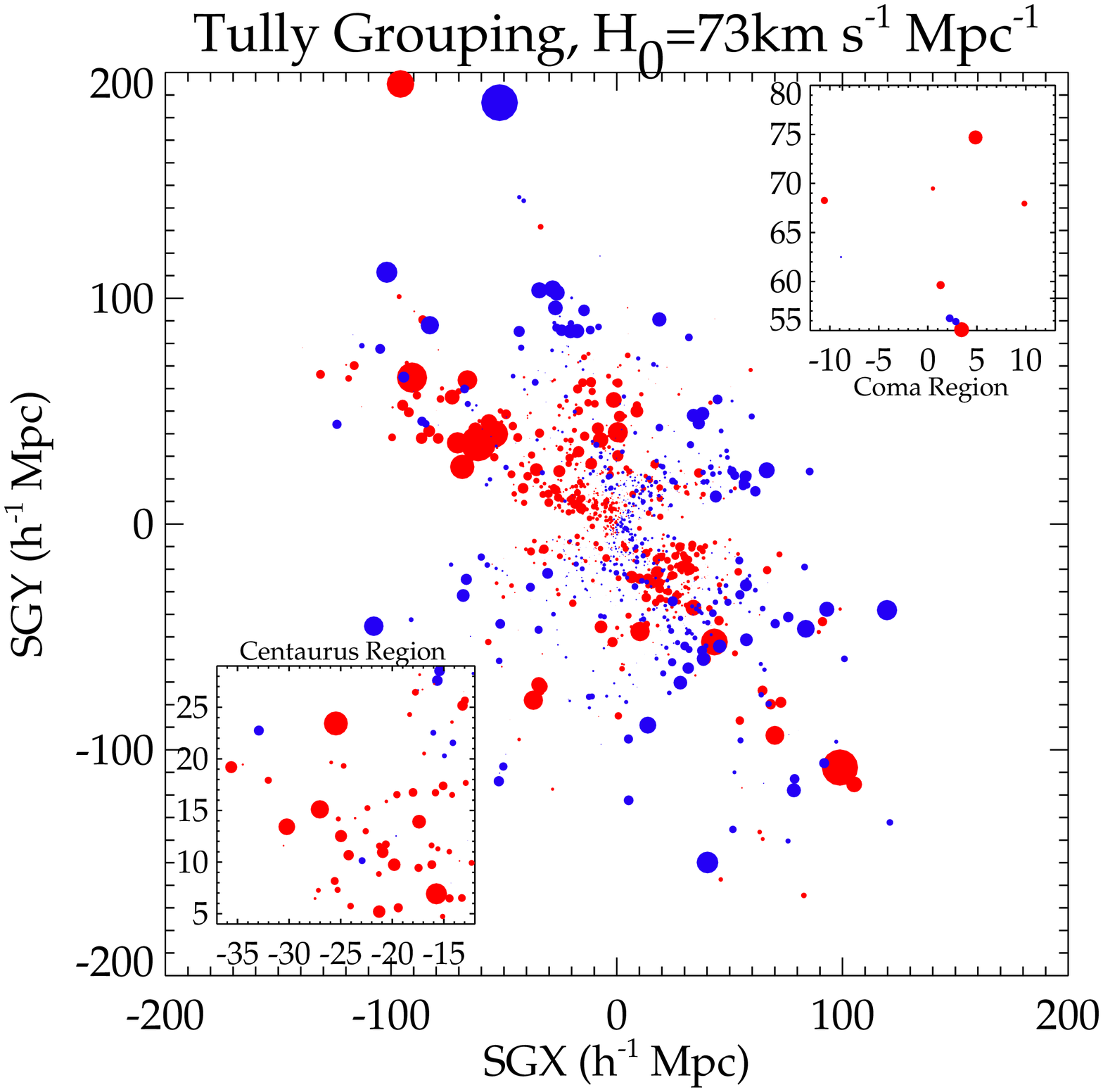}\\
\vspace{-1.5cm}

\hspace{-0.8cm}\includegraphics[width=0.55 \textwidth]{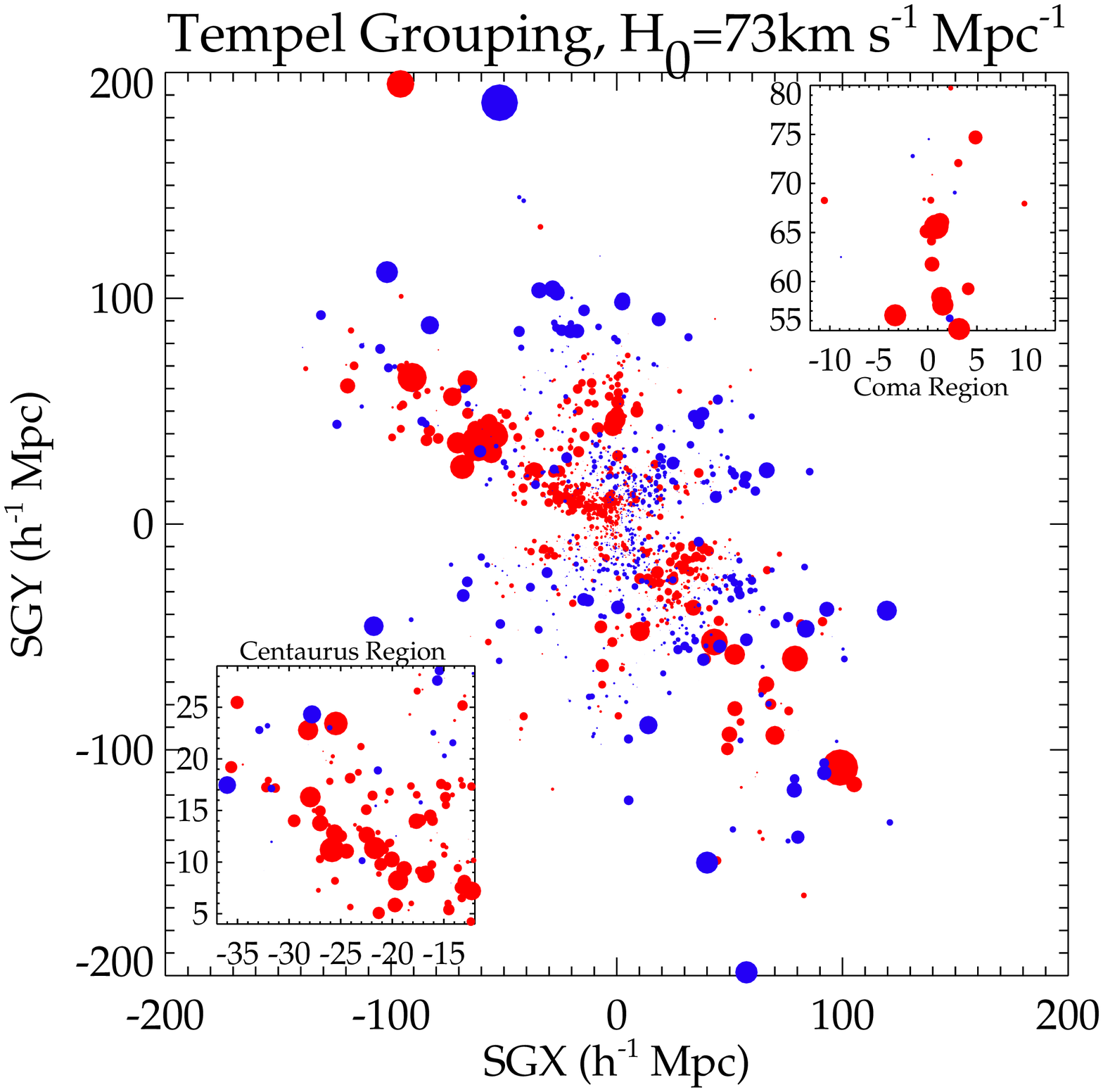}
\vspace{-1.5cm}

\caption{XY supergalactic slice (10~\hMpc) of the local Universe showing the constraints (radial peculiar velocity at galaxies' position) obtained with Tully (top) and Tempel (bottom) grouping schemes. A red dot means that the radial peculiar velocity is positive while a blue dot means that it is negative. The dot size is proportional to the absolute value of the radial peculiar velocity. Overall, the two grouping scheme exhibits catalogs in agreement with each other, the constraints are quite similar. However, zooming on a particularly dense region, like the Coma cluster area or the Centaurus cluster region, differences are more pronounced. Tempel grouping scheme presents more constraints with large values reinforcing the infall onto Coma/Centaurus (from both sides) that explains the contrast between Coma/Centaurus areas reconstructed using the second catalog of Cosmicflows obtained with the two different grouping schemes.}
\label{fig:tempeltullydata}
\end{figure}

Figure \ref{fig:tempeltullyfield} shows the residual between two WF reconstructions obtained with a different grouping scheme but with the same H$_0$ value. The figure is clear and irrevocable: while the velocity field is weakly affected by a different grouping scheme, the density field is largely impacted. Note that the position of structures is not impacted, structures are reconstructed at the proper location in both grouping scheme cases but their density value varies. In other words the infall onto the large structures is slightly more important with Tempel Grouping scheme than with Tully Grouping scheme.

Actually, the last third of Table~\ref{Tbl:2} gives the standard deviation of the residual velocity and overdensity fields of two reconstructions obtained with the exact same H$_0$ value but different grouping schemes. These values confirm that the velocity fields are quasi-identical except for the largest value of H$_0$ (76~\kms~Mpc$^{-1}$) but this value has been shown to be slightly unrealistic. This is in agreement with the fact that at larger H$_0$ values, Tully Grouping scheme produces overall a larger infall than Tempel Grouping scheme with the exception of H$_0$=75~\kms~Mpc$^{-1}$. Interestingly the standard deviation of the residual overdensity fields is not exceptionally high although Figure \ref{fig:tempeltullyfield} clearly shows that the structures are affected by the grouping schemes. The answer is in the maximum value of the residual overdensity fields. The standard deviation might be quite low but the maximum value is higher than when comparing for example two reconstructions obtained with different H$_0$ values but with the same grouping scheme. Consequently, on average the values of the residual are higher for reconstructions obtained with different grouping schemes than with different H$_0$ values confirming the observations made looking at the right panels of Figures \ref{fig:tullyfield} and \ref{fig:tempelfield} and at Figure \ref{fig:tempeltullyfield}.

To understand the difference emanating from the two grouping schemes in more detail, we look at the distribution of constraints in the XY supergalactic  
slice of the local Universe. Figure \ref{fig:tempeltullydata} shows the constraints as dots at galaxies' position: a blue dot means a radial peculiar velocity pointing towards us while a red dot stands for a radial peculiar velocity going away from us. The dot sizes are proportional to the radial peculiar velocity value in absolute value. In a first approximation, i.e. on large scales, the distributions of constraints and their values look overall very similar. Next, we focus on particular regions of interest such as the Coma cluster area that has been shown to present a structure with a greater contrast using Tempel Grouping rather than Tully Grouping. This particular region is plotted in the small top right inset in both panels of Figure \ref{fig:tempeltullydata}. The differences are striking: an infall from both sides of the Coma cluster region (red and blue) is visible in Tempel Grouped catalog while the more aggressive grouping used by Tully removes most of these constraints. The same goes for the Centaurus cluster region visible in the small bottom left inset in both panels of the same Figure. Tests made using different linking lengths in Tempel Grouping scheme confirm that a lesser grouping increases the infall/outflow and thus the contrasts of structures. This highlights the importance of a balance between grouping and removing non linear motions. 

\subsection{H$_0$: Not a real dependence}

\begin{figure}
\vspace{-2.5cm}

\hspace{-0.5cm} \includegraphics[width=0.58 \textwidth]{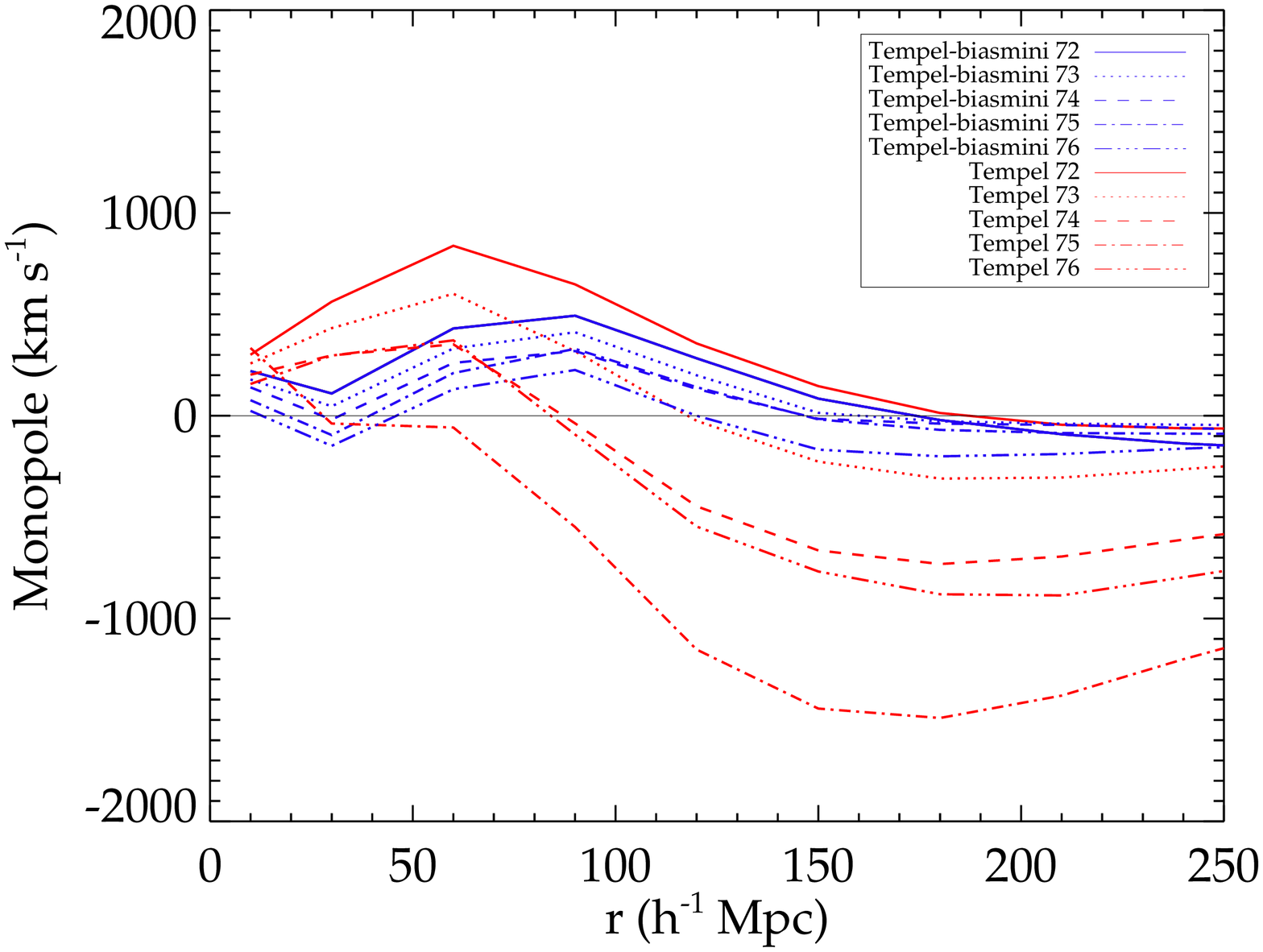}
 \vspace{-1.7cm}

\caption{Monopole of the Wiener Filter reconstructed fields as a function of the distance from us. The Wiener Filter has been applied to catalogs with different H$_0$ values (linestyle) using a unique grouping scheme but applying (blue) or not (red) a method to minimize the observational biases. H$_0$ impacts strongly the monopole term only for the catalogs without minimization of biases. The larger H$_0$, the larger is the infall on the local Volume. The minimization bias scheme has a strong influence on the monopole of the velocity field: it clearly suppresses the infall for all the values of H$_0$ considered. }
\label{fig:dipomonocorr}
\end{figure}

In this last part, we investigate whether the WF reconstruction has a real strong dependence on the Hubble constant value. Indeed, in the above tests, the bias minimization scheme developed originally to suppress the infall observed in the reconstructions has not been applied to the observational catalog. However, to build adequate constrained initial conditions, the observational catalog must undergo a bias minimization.  We apply the method described in \citet{2015MNRAS.450.2644S} to the different H$_0$ value catalogs grouped with Tempel scheme and run the WF technique on each one of them. Results are visible on Figure \ref{fig:dipomonocorr} in form of the monopole of the velocity fields. The bias minimization scheme strongly reduces the effect of the H$_0$ value selected to derive the peculiar velocities. There is clearly no strong infall anymore onto the local volume (no large negative values for the monopole at large radii) whatever H$_0$ value is used. This observation drastically  minimizes the previous conclusions about the dependence of the reconstruction on H$_0$ and removes concerns about choosing adequately H$_0$ providing that the catalog is bias minimized. 


\section{Conclusion}

Reconstructions of the three dimensional velocity and density fields of the local Universe are essential to study the local Large Scale Structure. Numerous methodologies have been developed to perform such reconstructions using observational data. In this paper, we use the Wiener Filter technique applied to galaxy radial peculiar velocity catalogs to obtain reconstructed velocity and overdensity fields of the local Volume. These reconstructions are useful as such for direct study of the linear local Universe today but also to build constrained initial conditions that permit performing constrained simulations of the local Universe, i.e. simulations that resemble the local Universe down to the cluster scales. We seek to understand how the Hubble constant value chosen to derive the radial peculiar velocities from galaxy distance measurements and total velocities, and the grouping scheme used to remove non linear motions affect the reconstructions and by extension impact the quality of the constrained simulations. 

To this end, two different grouping schemes (Tully based on the literature and Tempel based on a systematic algorithm) are selected as well as 5 reasonable locally derived H$_0$ values \citep[from 72 to 76~\kms~Mpc$^{-1}$, for the most recent values see e.g.][]{2016JCAP...08..026S,2016ApJ...826...56R,2016AJ....152...50T,2016ApJ...832..210B}. 10 grouped versions of the second radial peculiar velocity catalog of Cosmicflows are produced accordingly: 5 per grouping scheme with each one of the H$_0$ values. These catalogs differ by the number of isolated galaxies and groups as well as by their radial peculiar velocity distribution. Tully Grouping scheme results in more isolated galaxies but less groups and as a result less peculiar velocity-constraints when compared to Tempel Grouping scheme. Namely, the latter is found to be more aggressive than the former. In addition, the larger H$_0$, the more asymmetric the distribution, the larger the standard deviation and the more negative the mean.  

The WF algorithm is applied to these 10 catalogs and the resulting velocity and overdensity fields are compared. Whatever grouping scheme is used, the larger H$_0$ is, the larger the infall onto the local Volume is. If the tidal part of the velocity field due to objects outside of the local Volume is greatly affected by H$_0$, the divergent part due to the objects inside the Volume and tightly tied to the overdensity field is weakly impacted by a change in H$_0$. Note that it is the latter that is used to build constrained initial conditions. Actually, the latter is greatly affected by the grouping scheme. Comparing at fixed H$_0$, reconstructions obtained with catalogs grouped with different grouping schemes, we observe that structures, although they are present at the proper location in both cases, are more contrasted in Tempel grouping scheme's case then in Tully grouping scheme's case. This is in particular true for the Coma cluster area and the Centaurus cluster region. Looking for the reasons of such observations, we compare the distribution of radial peculiar velocity in the XY supergalactic slice and notice that overall the agreement between the catalogs grouped with the two different schemes is very good: positions of constraints (peculiar velocities) and their values match quite closely. However, when focusing on smaller areas to study the details, like the Coma cluster region or the Centaurus cluster region, we note quite a lot of differences mostly due to the difference in terms of aggressiveness of the grouping schemes. Tempel Grouping scheme allows more constraints in these regions than Tully Grouping scheme. Consequently the infall from both sides onto these areas are reinforced providing an explanation for the greater overdensity value in the reconstruction obtained with the Tempel grouped catalog than in that obtained with the Tully grouped catalog. Such findings highlight the importance of a balance between grouping to remove non linear motions and preserving some constraints to produce an infall onto structures that are expected to be large overdensities.

The main conclusions of the paper are as follows. The choice of H$_0$ impacts overall the velocity field in a given direction, i.e. it creates a general infall/outflow patterns but it does not really affect the overdensity field. Namely, the tidal part of the velocity field changes quite a lot with H$_0$ but not the divergent part. However, this conclusion has to be strongly mitigated. Indeed, the bias minimization scheme described in \citet{2015MNRAS.450.2644S} applied to the grouped observational catalog strongly suppresses the dependence of the reconstructions on H$_0$. There is no more drastic infall onto the local volume.  On the contrary, the grouping scheme affects greatly the overdensity field accentuating or diminishing the contrast between the structures. Still overall structures are reconstructed at the proper location with both grouping schemes studied here. Then in terms of H$_0$, we simply recommend either to choose the value giving the more neutral result (i.e. monopole term close to zero at large radii) or to apply the bias minimization scheme described in \citet{2015MNRAS.450.2644S} after grouping. Note that this bias minimization scheme also erases the bump entirely due to biases in the dipole term and makes the radial peculiar velocity distribution Gaussian. It is worth noticing that again, the dipole at large radii is proven to be very stable whatever choices is made to built the WF reconstruction providing that the grouping is properly done and that the catalog contains both clusters/groups and galaxies in the field \citep[e.g.][]{2017MNRAS.468.1812S}. Regarding the grouping scheme, there is a clear need for a balance between grouping to remove non linear motions to preserve the quality of the WF reconstruction \citep{2017MNRAS.468.1812S} and its stability with respect to H$_0$ choice, and keeping some constraints to contrast the high overdensity regions with respect to other regions. If a more aggressive grouping like Tully Grouping scheme permits stabilizing the WF reconstruction across a large range of H$_0$ values, preserving more constraints like with Tempel Grouping scheme provides Coma and Centaurus regions with a greater contrast with respect to other regions. Such an observation is promising to perform constrained simulations of greater quality than those of the first generation in terms of cluster (Virgo excluded since its mass is already in good agreement with observations) masses. Therefore, the next step consists in using the catalog grouped with Tempel Grouping scheme as constraints in order to build constrained initial conditions of the local Universe with the local massive clusters.

\section*{Acknowledgements}
We thank Brent Tully and Noam Libeskind for useful discussions as well as Yehuda Hoffman and Romain Graziani for interesting discussions and Stefan Gottl\"ober for useful comments. We acknowledge the use of the Extragalactic Distance Database to extract the Cosmicflows catalogs and thank the Cosmicflows team led by Brent Tully and Helene Courtois for the release of the catalogs. JS acknowledges support from the Astronomy ESFRI and Research Infrastructure Cluster ASTERICS project, funded by the European Commission under the Horizon 2020 Programme (GA 653477). ET acknowledges support from the ETAg grant IUT40-2 and from the European Regional Development Fund. We thank the anonymous referee for his/her comments.

\section*{Appendix A}
The Wiener Filter technique is the optimal minimal variance estimator given a dataset and an assumed prior power spectrum. Data dominate the reconstruction in region where they are dense and accurate. On the opposite when they are noisy and sparse, the reconstruction is a prediction based on the assumed prior model. Briefly, the overdensity $\delta^{WF}$ and velocity $\textbf{v}^{WF}$ fields of the Wiener Filter are expressed in terms of the following correlation matrixes. For a list of M constraints $c_i$:

\begin{equation}
\delta^{WF}(\textbf{r})=\sum_{i=1}^M \langle\delta(\textbf{r})c_i\rangle\eta_i,
\label{eq1}
\end{equation}

\begin{equation}
 v_{\alpha}^{WF}=\sum_{i=1}^M \langle v_{\alpha}(\textbf{r})c_i\rangle \eta_i \quad \mathrm{with} \quad \alpha=x,y,z,
 \label{eq2}
 \end{equation}

\noindent 
where $\eta_i=\sum_{j=1}^M\langle C_i C_j\rangle^{-1}C_j$ are the components of the correlation vector $\eta$. $C_i=c_i+\epsilon_i$ are observational constraints plus their uncertainties. Hence, $\langle C_i C_j\rangle$ is equal to $\langle c_i c_j\rangle+\epsilon_i^2\delta_{ij}$ assuming statistically independent errors. The constraints can be either densities or velocities. $\langle AB \rangle$ notations stand for the correlation functions involving the assumed prior power spectrum.\\
 
The associated correlation functions are given by: 
\[ \langle \delta(\textbf{r}\, ') v_{\alpha} (\textbf{r} \,'+\textbf{r}) \rangle  = \frac{\dot a f}{(2 \pi)^3}\int_0^\infty \frac{ik_{\alpha}}{k^2}P(\textbf{k}) e^{-i\textbf{k}.\textbf{r}}d \textbf{k} \]
\begin{equation} \; = -\dot a f r_{\alpha} \zeta (r) \end{equation}

\[  \langle v_{\alpha}(\textbf{r} \,')v_{\beta}(\textbf{r}\, '+\textbf{r})\rangle \;\; = \frac{(\dot a f)^2}{(2\pi)^3}\int_0^\infty \frac{k_{\alpha}k_{\beta}}{k^4}P(\textbf{k}) e^{-i \textbf{k} .\textbf{r}} d \textbf{k}  \]
\begin{equation} = (\dot a f)^2 \Psi_{\alpha\beta} \end{equation}

\noindent
where $P$ is the assumed prior power spectrum, $a$ the scale factor and $f$ the growth rate.\\

Because data sample a typical realization of the prior model, i.e. the power spectrum, $\frac{\chi^2}{d.o.f}$ should be close to 1 where $\chi^2=\sum_{i=1}^M\sum_{j=1}^M C_i\langle C_i C_j \rangle^{-1} C_j$ and d.o.f is the degree of freedom. However, data include non-linearities which are not taken into account in the model. Consequently, a  non linear sigma ($\sigma_{NL}$) such that $\langle C_i C_j \rangle = \langle c_i c_j \rangle+\delta_{ij}^k\epsilon_j^2+\delta_{ij}^k\sigma_{NL}^2$ is required to compensate for the non-linearities to drive $\frac{\chi^2}{d.o.f}$ closer to 1.\\


\bibliographystyle{mnras}

\bibliography{biblicomplete}
 \label{lastpage}
\end{document}